\documentstyle[aps,epsf,epsfig,twocolumn,psfig,array]{revtex}

\begin{document}

\draft \preprint{}

\title{Detecting $N$-particle interference patterns with linear balanced $N$-port analyzers}
\author{O. Steuernagel}
\address{Dept. of Physical Sciences,
University of Hertfordshire, College Lane, Hatfield, AL10 9AB, UK}
\date{\today}
\maketitle
\begin{abstract}
A standard two-path interferometer fed into a linear $N$-port
analyzer with coincidence detection of its output ports is
analyzed. The $N$-port is assumed to be implemented as a discrete
Fourier transformation $\cal F$, i.e., to be balanced. For unbound
bosons it allows us to detect $N$-particle interference patterns
with an $N$-fold reduction of the observed de Broglie wavelength,
perfect visibility and minimal noise. Because the scheme involves
heavy filtering a lot of the signal is lost, yet, it is
surprisingly robust against common experimental imperfections, and
can be implemented with current technology.
\end{abstract}

\narrowtext

%\section{Introduction}\label{introduction}

A quantum state containing $N$ particles can yield an $N$-fold
reduction of the observed de Broglie wavelength in an interference
experiment when compared with a single particle
pattern~\cite{Jacobson95}. This can lead to an $N$-fold increase
of interferometric sensitivity, in the best case, reaching the
quantum mechanical 'Heisenberg-limit' of particle-number--based
enhancement in resolution~\cite{Ou96}. For bound particles this is
straightforward~\cite{Arndt99}, as long as they are not split up
by the interferometer, namely, their binding energy exceeds the
interaction energies encountered when passing the interferometer's
beam splitters and mergers~\cite{Jacobson95}. But, it is difficult
to reach the Heisenberg-limit, or, more generally speaking, to
detect $N$-particle patterns in imaging or interferometry using
{\em unbound} particles~\cite{Caves81,Burnett93,{Huelga97}} such
as free photons.

Experimental results have so far only demonstrated a halving of
the observed wavelength using two-photon
states~\cite{Fonseca99,{dAngelo},{Edamatsu02}} or reported
signatures of four-particle entanglement for
photons~\cite{Bouwmeester,Ou99} and ions~\cite{Sackett00}. Yet,
weakly or unbound bosons such as photons or cold atoms are
currently the most important particles for interferometry and
imaging. Interest in several recent schemes to employ their
multi-particle features in
inter\-fero\-meters~\cite{Jacobson95,Caves81,Burnett93,Fonseca99,dAngelo,Edamatsu02,Ole.02.deBroglie},
high resolution imaging~\cite{kolobov00}, and quantum
lithography~\cite{lithography} has therefore been
considerable~\cite{NewScientist01}.

Observing multi-particle quantum effects for unbound particles is
hampered by several problems: special quantum states are needed
that are hard to
synthesize~\cite{Jacobson95,Ou96,Burnett93,{kolobov00},lithography},
the processing involves large non-linear particle-particle
interactions~\cite{Jacobson95} or joint detection of all particles
with single particle
resolution~\cite{Burnett93,Fonseca99,dAngelo,Edamatsu02,Ole.02.deBroglie,lithography}.

\begin{figure}
\epsfverbosetrue \epsfxsize=3.2in \epsfysize=1.9in
%
%\epsffile{Setup.eps}
\epsffile[076 544 552 715]{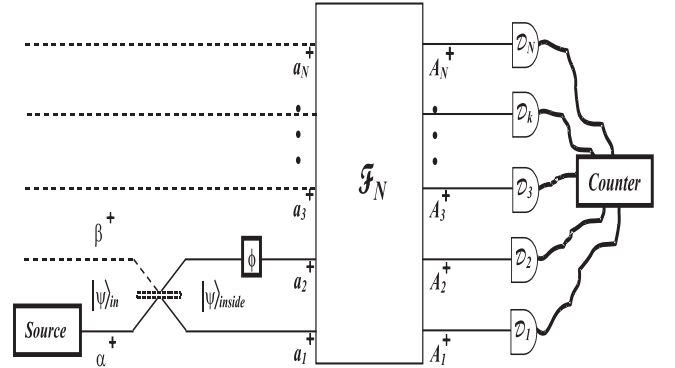}
\caption{Sketch of the setup: dotted lines depict $N-2$ empty
(vacuum) modes fed into the Fourier mixer ${\cal F}_N$ together
with the two active interferometer modes $a^\dagger_1$ and
$a^\dagger_2$ resulting from mixing the modes $\alpha^\dagger$ and
$\beta^\dagger$ at a balanced beam splitter. Mode $a^\dagger_2$
suffers the interferometric phase shift $\phi$. The corresponding
output modes of the Fourier mixer ${\cal F}_N$ are $A^\dagger_1$
to $A^\dagger_N$; they are detected by $N$ single-photon detectors
$\cal D$ which are read out in coincidence. \label{figure1}}
\end{figure}

The {\em linear} scheme presented in this paper shows that
equipping an interferometer's output with a balanced $N$-port
analyzer allows us to circumvent several of the above problems and
detect multi-photon interference patterns with current technology.
Note, that other linear schemes have recently been devised to
circumvent similar problems in quantum information
theory~\cite{Milburn.Nat,Zeilinger.Nat} and quantum state
preparation~\cite{ole.97.oc,{Lee02}}. Using the scheme presented
here it should be straightforward to observe, for the first time,
a more than two-fold reduction of the effective de Broglie
wavelength of unbound
photons~\cite{Jacobson95,Fonseca99,dAngelo,Edamatsu02,Ole.02.deBroglie};
the scheme is sketched in FIG.~\ref{figure1}.
%
%%
%%%
%%%
%%
%
In order to model our system let us follow the evolution of a state of
bosonic particles starting out in channel $\alpha^\dagger$ of
FIG.~\ref{figure1} and progressing from left to right. For the
transformations describing the mixing of the inter\-fero\-meter's input
modes by a balanced beam-splitter in conjunction with the phase shift $\phi$
for mode $a^\dagger_2$ we choose the following operator equations
%
%%
%%%
\begin{eqnarray}
\hat a_1^\dagger = ({\hat \alpha^\dagger + \hat \beta^\dagger})/{\sqrt 2}, \quad \hat
a_2^\dagger = e^{-i\phi}({\hat \alpha^\dagger - \hat \beta^\dagger})/{\sqrt 2} \; .
 \label{beam.splitter.mode.transformation}
\end{eqnarray}
%%%
%%
%
Subsequent to the interferometer the balanced $N$-port ${\cal
F}_N$ acts as a discrete Fourier-transformer~\cite{Reck94,Paul96}
and mixes the interferometric modes $a^\dagger_1$ and
$a^\dagger_2$ with the $N-2$ vacuum modes
$a^\dagger_3,\ldots,a^\dagger_N$. Using a vector notation for
these input modes $ a_k^\dagger$ and the output modes $A_j^\dag$,
i.e. $\vec{\hat{A}^\dag}={\cal F}_N \; \vec{\hat{a}^\dag}$, we
specify the corresponding (Fourier-) transformation matrix
elements as
%
%%
%%%
\begin{eqnarray}
\left( {\cal F}_N\right)_{j\; k} = {\frac {1}{\sqrt {N}}} \;
\exp{[ i \frac{ 2 \pi }{N} {\left (j-1\right )\left (k-1\right )}
] }
 \; . \label{F.N.port.matrix}
\end{eqnarray}
%%%}}
%%
%
To make a connection with the textbook case of classical or single
photon interference~\cite{ole.PRA.01.classical} let us look at the
case $N=2$, the conventional Mach-Zehnder inter\-fero\-meter. For
the input state we assume a single photon is entering through
channel $\alpha^\dag$, namely $|\psi \rangle_{in} = \hat
\alpha^\dag | 0 \rangle$. After the
transformations~(\ref{beam.splitter.mode.transformation})
and~(\ref{F.N.port.matrix}) this state is converted into $|\psi
\rangle_{in} = (1+e^{-i\phi})/2\;(\hat A_1^\dag - \hat A_2^\dag) |
0 \rangle$ yielding the customary classical interference pattern
$\langle \hat I_1 \rangle = \langle \hat A_1^\dag \hat A_1 \rangle
= (1+\cos\phi)/2 $ and the antifringe pattern in the second
channel $\langle \hat I_2 \rangle = (1-\cos\phi)/2 \,$. Of course,
the coincidence pattern $\langle \hat I_1 \cdot \hat I_2 \rangle =
0 $ vanishes, since only a single photon is present. The simplest
non-trivial case is that of the Mach-Zehnder interferometer fed
with a two-photon state $ |\psi_{(2)} \rangle_{in} = \frac{ ( \hat
\alpha^\dag )^2 }{\sqrt 2} | 0 \rangle$: it yields the
classical~\cite{ole.PRA.01.classical} first order interference
patterns $\langle \hat I_{1/2} \rangle = 1 \pm \cos \phi$ and the
coincidence signal $\langle \hat I_1 \cdot \hat I_2 \rangle =
(1-\cos 2 \phi)/4\, $, which shows the desired halving of the
effective de Broglie wavelength and full
contrast~\cite{ole.PRA.01.classical} observed in
experiments~\cite{Fonseca99,dAngelo,Edamatsu02}.

Generalizing the above discussion we now consider the $N$-{\em
channel coincidence count operator} $\hat {\cal I}_N$
%
%%
%%%
\begin{eqnarray}
 \hat {\cal I}_N  & \doteq & \prod_{j=1}^{N} \hat {I}_j  =
 \prod_{j=1}^{N} \hat A_j^\dagger \cdot \prod_{k=1}^{N} \hat {A}_k
\;,
  \label{NN.operator}
\end{eqnarray}
%%%}}
%%
%
where we have used the fact the the output modes $A$ commute.
Expressing $\prod_{k=1}^{N} \hat {A}_k $ in terms of the input
operators $\hat a^\dagger_1$ and $\hat a^\dagger_2$ yields
%
%%
%%%
\begin{eqnarray}
 & \hat {\cal I}_N =
 \frac{  \hat a_1^{\dag \, N} - (- \hat a_2^{\dag})^N  }{ \sqrt N^N} \cdot
 \frac{   \hat a_1^{ \, N} - (- \hat a_2^{})^N }{ \sqrt N^N}
  \label{NN.operator.1b} \\
%& = \frac{1}{ N^N}  \left( \hat a_1^{\dag \, N} \hat a_1^{ \, N} + \hat a_2^{\dag \, N} \hat a_2^{N} - (-1)^N \left[ \hat a_1^{\dag \, N} \hat a_2^{N} + \hat a_1^{ \, N}  \hat a_2^{\dag \, N} \right] \right) \\
& =  \frac{\left( \hat a_1^{\dag \, N} \hat a_1^{ \, N} +  \hat
a_2^{\dag \, N} \hat a_2^{N} - (-1)^N \left[ \hat a_1^{\dag \, N}
\hat a_2^{N} + \hat a_1^{ \, N}  \hat a_2^{\dag \, N} \right]
\right)
 }{ N^N}  \; , \label{NN.operator.2}
\end{eqnarray}
%%%}}
%%
%
where we have used~(\ref{F.N.port.matrix}) and assumed that $\hat
a^\dagger_3 = \hat a^\dagger_4 = ... = \hat a^\dagger_N = 0$ in
accordance with our scheme's stipulation to leave these modes
empty (tracing out the vacuum state).

Note, that the terms $\hat a_j^{\dag \, N} \hat a_j^{ \, N} $
represent generalized $N$-particle intensity or $N$-particle
dosage terms~\cite{lithography}. Correspondingly, the terms in
square brackets represent $N$-particle cross-mode terms which
generalize the well known single particle (or classical)
interference terms~\cite{ole.PRA.01.classical} of the form $\hat
a_1^{\dag } \hat a_2 + \hat a_1 \hat a_2^{\dag } $.

Eq.~(\ref{NN.operator.2}) contains a good and a bad message, the
bad aspect is the emergence of the {\em signal suppression factor}
$N^{-N}$ which is due to the fact that we require all output
detectors to fire -- a rare event: $\langle \hat {\cal I}_N
\rangle =0 $ if one of the output channel detectors fails to fire.
Such losses are common whenever one tries to substitute non-linear
by linear
elements~\cite{Milburn.Nat,{Zeilinger.Nat},{ole.97.oc},{Lee02}}.

The positive aspect is the result that the form of $\hat {\cal
I}_N$ in Eq.~(\ref{NN.operator.2}) suggests that for those
subensembles of events that trigger firing of all $N$ detectors a
{\em 'perfect' $N^\textrm{th}$ order interference pattern
measurement} can be performed. Only linear elements and an array
of conventional detectors is needed for its implementation, it
will turn out that the scheme is rather robust with regards to
detector imperfections. Current technology suffices,
Eq.~(\ref{NN.operator.2}) therefore is important, it is the main
result of this paper.

In order to prove that $ \hat {\cal I}_N $ does indeed describe a
'perfect' $N^\textrm{th}$ order interference pattern measurement
let us check its signal and noise properties. The combination of
operators suggests that $ \hat {\cal I}_N $ probes for the
presence of the number-entangled state
%
%%
%%%
\begin{eqnarray}
| \psi \rangle_{inside} = \frac{| N \rangle_{a_1} | 0
\rangle_{a_2} + e^{iN\phi} | 0 \rangle_{a_1} | N \rangle_{a_2}
}{\sqrt{2}} \; . \label{best.state.inside}
\end{eqnarray}
%%%
%%
%
This is the superposition-state needed when trying to reach the
Heisenberg limit in interferometry\cite{Ou96} using single photon
count techniques. Note, that to synthesize this kind of state one
typically needs correlated input in both modes $\hat
\alpha^\dagger$ and $\hat \beta^\dagger$, see e.g.~\cite{Lee02},
this case is obviously not covered by our sketch of the setup in
FIG.~\ref{figure1} where the source is only seen to feed into
mode~$\hat \alpha^\dagger$.

Its signal is $\langle \hat{\cal I}_N \rangle = N!/N^N \times
(1-(-1)^N \cos (N \phi))$ and does indeed display the term $\cos
(N \phi)$ due to an $N$-fold decreased effective de Broglie
wavelength~\cite{Jacobson95,Fonseca99,dAngelo,Edamatsu02,Ole.02.deBroglie}
and an {\em interference pattern with perfect visibility}.

The second moment is $\langle \hat{\cal I}_N^2 \rangle =
N!^2/N^{2N} \sin^2 (N \phi)$.

To study the scheme's noise features, we will estimate the noise
induced phase spread $\Delta\phi$ from the quotient of the
signal's fluctuations and the signal's phase gradient (see a good
textbook such as~\cite{Scully.buch})
%
%%
%%%
\begin{eqnarray}
\Delta\phi = \Delta {\cal I}_N / | \partial {\cal I}_N /
\partial \phi|
%\mbox{, where } \Delta P_N = \sqrt{ \langle P_N^2 \rangle - \langle P_N \rangle^2 }
 \; , \label{phase.resolution}
\end{eqnarray}
%%%}}
%%
%
with $\Delta {\cal I}_N = \sqrt{ \langle \hat{\cal I}_N^2 \rangle
- \langle \hat{\cal I}_N \rangle^2 }$. For the number-entangled
state~(\ref{best.state.inside}) the phase spread therefore is
%
%%
%%%
\begin{eqnarray}
\Delta\phi = 1/N \; . \label{1overN}
\end{eqnarray}
%%%
%%
%
The subensemble of registered events (all detectors fire) reaches
the Heisenberg-limit~\cite{Scully.buch} which proves that our
scheme has {\em the least possible noise}~\cite{Ou96}.

According to Ou's analysis~\cite{Ou96} this limit cannot be
reached using arbitrary states, i.e. in our case, states other
than~(\ref{best.state.inside}).

Let us therefore study the signal and noise properties of some
representative states that are fed through one channel, e.g.
$\alpha^\dagger$, only. To start out, let us consider $N$-photon
Fock states $ |\psi_{(N)} \rangle_{in} \doteq \frac{ ( \hat
\alpha^\dag )^N }{\sqrt N!} | 0 \rangle$, which can in principle
be synthesized using projection measurements~\cite{ole.97.oc}; the
signal~$\langle \hat {\cal I}_N \rangle$ has the form
%
%%
%%%
\begin{eqnarray}
\langle \hat {\cal I}_N \rangle  \doteq \langle \prod_{j=1}^{N} \hat {I}_j \rangle  = {\frac
{N!}{2^{(N-1)} \, N^N}} \; \left ( 1 -(-1)^N \cos(N\phi) \right )
 \; . \label{NN.pattern}
\end{eqnarray}
%%%}}
%%
%
We see that an interference pattern~(\ref{NN.pattern}) with
$N$-fold reduction and perfect contrast is observable, yet we will
find that the imperfect matching to the ideal
state~(\ref{best.state.inside}) leads to large noise.

According to Stirling's formula, this signal scales like
$\frac{N!}{2^{(N-1)} \, N^N} $ $\approx$ $ {\frac {\sqrt {8\pi
\,N}}{\left (2\,{e}\right )^{N}}}$, with the corresponding
numerical values $1/4, $ $1/18, $ $3/256, $ $3/1250, $ $5/10368$
for $N=2,3,...,6$. Evidently, the requirement for simultaneous
firing of all detectors reduces the signal strength considerably.

The joint-detection operator $\hat {\cal I}_N$ extracts suitable
components to show $N$-photon interference patterns out of every
state with sufficiently large photon numbers. In order to prove
this for general input into channel $\alpha^\dagger$ let us now
consider an excess $E$ of particles $ N_{state} = N+E $ over the
number of ports $N$. Trivially, a negative excess, i.e. a number
of particles fewer than expected detector clicks will give a zero
signal. For a Fock-state $ |\psi_{(N+E)} \rangle_{in} \doteq
\frac{ ( \hat \alpha^\dag )^{N+E} }{\sqrt{ (N+E)!} } | 0 \rangle$
with a positive excess the corresponding intensity expression
shows the same pattern as Eq.~(\ref{NN.pattern}) above -- with
favorably increased magnitude
%
%%
%%%
\begin{eqnarray}
\langle \hat {\cal I}_N \rangle  = \frac {(N+E)!}{E!} \cdot \frac
{ 1 -(-1)^N \cos(N\phi) }{2^{(N-1)} \, N^N}
 \; . \label{NNEx.pattern}
\end{eqnarray}
%%%}}
%%
%
In order to appreciate the very considerable intensity gain
attributable to such a photon-number excess, note, that at the
rather modest cost of doubling the number of photons, one can
roughly compensate the losses due to an increased number of ports.
For example, the losses due to moving from $N=2$ photons and
channels to 10 photons and channels is offset by moving to 10
channels and $20$ photons instead. Moreover, large photon number
excess also reduces the noise, see below.

In order to generalize our results to states other than
Fock-states let us maintain that the $\beta$-channel is empty and
let us trace it out: $\beta = \beta^\dagger=0$. In this case the
coincidence count operator $ \hat {\cal I}_N$ commutes with the
input photon number operator $\widehat{\alpha}^\dag
\widehat{\alpha}$ implying that any Fock-state component exceeding
the port number $N$ contributes separately to the overall
interference pattern and its noise features. Hence, for a general
$\alpha$-channel input state $ | \psi \rangle_{in}  =
\sum_{J=0}^{\infty} c_J \frac{\hat \alpha^{\dag \, J} }{\sqrt{J!}}
| 0 \rangle , $ the coincidence pattern has the form
%
%%
%%%
\begin{eqnarray}
\langle \hat {\cal I}_N \rangle  = \sum_{E=0}^{\infty} |c_{N+E}|^2
\frac {(N+E)!}{E!} \cdot \frac { 1 -(-1)^N \cos(N\phi) }{2^{(N-1)}
\, N^N}
 \; . \label{NN.sum.pattern}
\end{eqnarray}
%%%}}
%%
%
A corresponding generalization applies to mixed input states; this
shows that our scheme allows for a robust formation of an $N$-fold
reduced interference pattern with perfect visibility in the
general case of states fed through channel $\alpha^\dagger$ only.

Let us now discuss the noise properties of the $N$-photon state $
|\psi_{(N)} \rangle_{in} = \frac{ ( \hat \alpha^\dag )^N }{\sqrt
N!} | 0 \rangle$. The second moment is $\langle \hat I_N^2 \rangle
= N!^2/(N^{2N} 2^{N-2}) $ $ \times (1-(-1)^N\cos(N\phi))$ yielding
the mini\-mum phase estimate (at positions $\phi=0$ for even $N$
and $\phi=\pi/N$ for odd $N$)
%
%%
%%%
\begin{eqnarray}
\Delta\phi = \frac{\sqrt{2^{N-1}}}{N} \; . \label{Noise.N}
\end{eqnarray}
%%%}}
%%
%
A Fock state fed into the $ \alpha^\dagger $-channel has little
overlap with the number-entangled state~(\ref{best.state.inside}).
This leads to extra noise increasing the phase spread by the
factor $\sqrt{2^{N-1}}$.

\begin{figure}
\epsfverbosetrue \epsfxsize=3.2in \epsfysize=2.0in
%
%
%\epsffile[000 100 592 666]{ldDeltaphi.2.eps}
\epsffile[000 158 520 570]{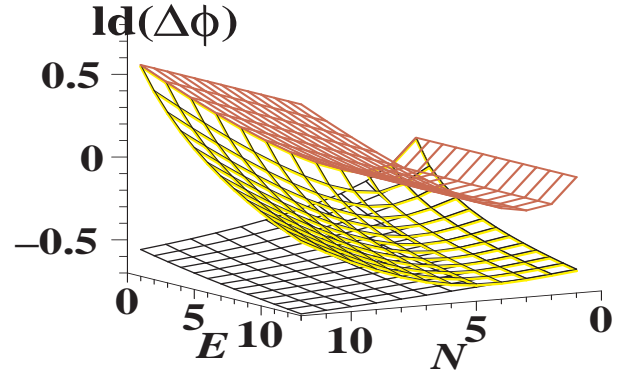}
\caption{Logarithmic plot of minimum phase spread $\Delta\phi$
(logarithm 'ld' to basis 10) for setups with $N$ ports. The poor
performance~\protect{(\ref{Noise.N})} of a Fock state with $N$
photons fed into the \protect{$ \alpha^\dagger $}-channel (top
sheet) is compared to shot noise $\Delta\phi=1/\sqrt{N+E}$ (bottom
sheet). Shot noise can be reached employing coherent states
$|\sqrt{N+E}\rangle_\alpha$. For a Fock-state $ |\psi_{(N+E)}
\rangle_{in} $ with sufficiently large photon excess $E$ over the
number of ports $N$ (middle sheet) the noise is reduced, reaching
(but not falling below) shot noise. \label{figure2}}
\end{figure}

We have seen that the use of Fock states with a photon number
excess $N+E$ over the port number $N$ leads to a considerable
signal increase~(\ref{NNEx.pattern}). Also the noise performance
in this case is better. The corresponding expression for the noise
is a bit involved (it was determined using the 'Maple' symbolic
manipulation program). For large photon number excess $E$ we find
that the noise falls to the shot-noise level, but not below; see
FIG.~\ref{figure2}.
%
%%
%%%
%%%
%%
%
Note, that we restricted most of the above analysis to the case of
states fed through the input channel~$\alpha^\dag$ only, this is
because there are few transparent analytical expressions in the
general case. However, in order to go below the shot noise level
quantum correlated input, using channel~$\beta^\dag$ as well, is
needed (remember Eq.~(\ref{1overN})); this can either reduce or
enhance the interferometric signal~\cite{ole.PRA.01.classical}.
Our discussion shows that currently available quantum states can
be used and that moving towards the two-mode entangled number
state~\cite{Lee02} allows us not only to observe an $N$-fold
reduced de Broglie wavelength but also noise below the shot-noise
level (for the subensembles selected by the coincidence clicks).

Our scheme is not only versatile with regards to the states that
can be used, it also is quite insensitive to detector
imperfections (other than dark counts) which often lead to serious
degradation for quantum state reconstruction schemes~\cite{ole95}.
Let us first consider detector losses, described by the admixture
of vacuum modes $V^\dagger_j$ into every detector mode $\hat
A^\dagger_j \mapsto \tau_j^* \hat A^\dagger_j + \rho_j^* \hat
V^\dagger_j $, where $|\tau_j|^2+|\rho_j|^2=1$. We find that the
multi-channel coincidence signal operator for lossy detectors
$\hat {\cal I}_{N, lossy}$ acquires the form
%
%%
%%%
\begin{eqnarray}
 \hat {\cal I}_{N, lossy}  & = & \prod_{j=1}^{N}
 (\tau_j^* \hat A^\dagger_j + \rho_j^* \hat V^\dagger_j )
 (\tau_j \hat A_j + \rho_j \hat V_j )
 \\
  & = &  \prod_{k=1}^{N} |\tau_k|^2 \prod_{j=1}^{N} \hat A^\dagger_j \hat A_j \doteq T \prod_{j=1}^{N} \hat I_j = T
 \hat {\cal I}_N
 \; . \label{NN.operator.robust}
\end{eqnarray}
%%%}}
%%
%
Again, the fact that all modes $V^\dagger_j$ are vacuum modes and can be
traced out by setting $ V_j=0$ has been used. The only net-effect of
detector losses is the rescaling of the signal strength by the transmission
loss factor $T \doteq \prod_{k=1}^{N} |\tau_k|^2$. Since this factor cancels
when forming the appropriate quotients of Eq.~(\ref{phase.resolution}) it
does not even affect the phase-resolution properties directly.

Most detectors are unable to distinguish one from two and more
photons, see~\cite{good.detectors} though. This inability often
poses problems~\cite{{Burnett93},Scully.buch,ole95} which our
present scheme is also quite insensitive to. This can be seen from
the following argument: if only one photon exits per channel $
A_j^\dagger$ there obviously is no difference in detection by
number-sensitive detectors and those that only discriminate
between presence and absence of photons. For the later case let
us, in analogy to expression~(\ref{NN.operator}) define the
N-channel {\em photon-presence} operator $\hat P_N =
\prod_{j=1}^{N} \left(\hat 1_j - |0\rangle_j {}_j\langle 0|
\right) $ where the projectors 'unity minus vacuum' $\left(\hat
1_j - |0\rangle_j {}_j\langle 0| \right)$ project out the vacuum
modes in the exit channels $ A_j^\dagger$. This is equivalent to
saying that we determine the signal of the $N$-channel coincidence
count operator $\hat {\cal I}_N$ divided by the respective photon
number in the output $\hat P_N = \hat P_N \hat {\cal
I}_N(\prod_{j=1}^{N} \hat A^\dagger_j \hat A_j )^{-1} \hat P_N$.
Consequently $\langle \hat P_N (\phi)\rangle \leq \langle \hat
{\cal I}_N (\phi)\rangle$ and $\langle \hat P_N (\phi)\rangle > 0$
iff $ \langle \hat {\cal I}_N (\phi)\rangle > 0$. In other words
$\langle \hat P_N (\phi)\rangle$ and $ \langle \hat {\cal I}_N
(\phi)\rangle$ have the same zeros and similar global behavior:
the same period in interference patterns can be seen. It is not
difficult to show that the diminished signal $\langle \hat P_N
(\phi)\rangle$ encounters increased noise though.

Note the superficial similarity of our setup to imaging in the
Fraunhofer-limit, see FIG.~\ref{figure1}. The 'object' (beam
splitter plus phase shifter) is being mapped by a 'lens'
($N$-port) into the 'far-field' (detector array $A_j^\dagger$). It
remains to be seen whether the discussion presented here can be
extended to quantum enhanced imaging~\cite{kolobov00}.

It should also be interesting to investigate our setup for
correlated input states feeding several of the input modes.

To conclude: we have studied an inter\-fero\-metric setup using an
$N$-port readout containing only linear elements. The $N$ output
detectors have to fire in coincidence, this way a lot of signal
and noise are filtered away. The selected subensemble can reach
the Heisenberg limit which proves that the scheme is
noise-optimized. The scheme is versatile with respect to the type
of quantum states that can be employed, allows us to see an
$N$-fold reduction of the measured de Broglie wavelength with
perfect visibility, is surprisingly insensitive to common detector
imperfections, and readily implementable using present technology.
\acknowledgments I wish to thank Antia Lamas-Linares, Koji
Murakawa, Susana Huelga, Tony Chefles, and John Vaccaro for lively
discussions.

\end{document}